\documentclass[twocolumn,jmp,10pt,superscriptaddress,aps,nofootinbib]{revtex4-1}

\usepackage{amsmath}
\usepackage{amssymb}
\usepackage{graphicx}
\usepackage{hyperref}
\usepackage{color}

\usepackage[T1]{fontenc}
\usepackage[utf8]{inputenc}

\usepackage{amsthm}
\usepackage{bbm}
\usepackage{mathtools}

\newtheorem{thm}{Theorem}[section]
\newtheorem{defi}[thm]{Definition}
\newtheorem{lem}[thm]{Lemma}
\newtheorem{prop}[thm]{Proposition}

\def\tr{\operatorname{tr}}
\def\idty{\mathbbm1} 
\def\id{{\rm id}}
\def\Rl{{\mathbb R}}\def\Cx{{\mathbb C}}
\def\Ir{{\mathbb Z}}\def\Nl{{\mathbb N}}
\def\Rt{{\mathbb Q}}

\def\norm #1{\Vert #1\Vert}

\def\ketbra #1#2{{\vert#1\rangle\langle#2\vert}}
\def\kettbra#1{\ketbra{#1}{#1}}

\def\tr{\mathop{\rm tr}\nolimits}

\def\dd{\mbox{d}}                                   

\def\BB{{\mathcal B}}\def\HH{{\mathcal H}}

\def\EF{E} 
\def\nr{p} 
\def\dr{q} 

\renewcommand{\d}{\ensuremath{d}}

\def\idty{{\leavevmode\rm 1\mkern -5.4mu I}} 
\def\id{{\rm id}}
\def\Rl{{\mathbb R}}\def\Cx{{\mathbb C}}
\def\Ir{{\mathbb Z}}\def\Nl{{\mathbb N}}
\def\Rt{{\mathbb Q}}
\def\norm #1{\Vert #1\Vert}

\def\ketbra #1#2{\vert #1\rangle \langle #2\vert}
\def\kettbra#1{\ketbra{#1}{#1}}

\def\magtcont#1{\widetilde P_{#1}}

\def\Ts{{\mathcal T}} 

\def\hol{{\rm Hol}}
\def\Tf{{T\mkern-12mu^{=}\mkern2mu}} 
\def\Tf{T^\flat}

\def\ggroup{{\mathcal G}}
\def\Tsf{\Ts^\flat}
\def\Tsff#1{\Ts^{\flat\,#1}}

\def\proofdir#1#2{\noindent{\it (#1)$\Rightarrow$(#2):\ }}

\usepackage{tikz}
\usetikzlibrary{shapes,arrows,positioning}

\begin{document}

\title{Quantum walks in external gauge fields}

\author{C. Cedzich}
\affiliation{Institut f\"ur Theoretische Physik, Leibniz Universit\"at Hannover, Appelstr. 2, 30167 Hannover, Germany}
\affiliation{Institut f\"ur Theoretische Physik, Universit\"at zu K\"oln, Z\"ulpicher Stra{\ss}e 77, 50937 K\"oln, Germany}
\author{T. Geib}
\affiliation{Institut f\"ur Theoretische Physik, Leibniz Universit\"at Hannover, Appelstr. 2, 30167 Hannover, Germany}
\author{A.~H. Werner}
\affiliation{{QMATH}, Department of Mathematical Sciences, University of Copenhagen, Universitetsparken 5, 2100 Copenhagen, Denmark,}
\affiliation{{NBIA}, Niels Bohr Institute, University of Copenhagen, Denmark}
\author{R. F. Werner}
\affiliation{Institut f\"ur Theoretische Physik, Leibniz Universit\"at Hannover, Appelstr. 2, 30167 Hannover, Germany}

\begin{abstract}
Describing a particle in an external electromagnetic field is a basic task of quantum mechanics. The standard scheme for this is known as ``minimal coupling'', and consists of replacing the momentum operators in the Hamiltonian by modified ones with an added vector potential. In lattice systems it is not so clear how to do this, because there is no continuous translation symmetry, and hence there are no momenta. Moreover, when time is also discrete, as in quantum walk systems, there is no Hamiltonian, only a unitary step operator. We present a unified framework of gauge theory for such discrete systems, keeping a close analogy to the continuum case. In particular, we show how to implement minimal coupling in a way that automatically guarantees unitary dynamics. The scheme works in any lattice dimension, for any number of internal degree of freedom, for walks that allow jumps to a finite neighborhood rather than to nearest neighbours, is naturally gauge invariant, and prepares possible extensions to non-abelian gauge groups.
\end{abstract}

\maketitle

\section{Introduction}
In textbook quantum mechanics particles are placed in an electromagnetic field by applying the so-called \emph{minimal coupling principle}. This demands a modification of the generators of space and time translations, such that they only commute with each other up to multiplication operators, which reflects the presence of electromagnetic fields.

In this manuscript the systems under consideration are single particles with spin which evolve in discrete time on a lattice. Such systems are called \emph{quantum walks} \cite{Ambainis2001,Grimmet,SpaceTimeCoinFlux,TRcoin,GenMeasuringDevice}. The minimal coupling mechanism for turning on an  electromagnetic field does not carry over to such systems directly, because discrete translations have no  \emph{generators} $P_\mu$. However, one would expect an analogue to hold, where the self-adjoint generators are replaced by unitary one-step translation operators, both in space and in time, and no reference to a background field on a space-time continuum is needed. The first main message of this paper is that this works, and we set up the necessary lattice gauge environment.  The important second message is how to set up the minimal coupling scheme, that is, how to put a given walk into an external field, and specifically, what to substitute in the magnetic substitution and how. Of course, our scheme is gauge invariant in the sense that, up to a gauge transformation, the result depends only on the field, and  not on the vector potentials. While there are several examples in the literature \cite{ewalks,electric2D,othermagwalk,Sajid,Farrelly15} which describe an external electromagnetic field in accordance with our general scheme, the general case has not been formulated and studied at its natural level of generality. Hence we give a unifying background to the works mentioned, and also help to correct some faulty approaches [no cite].

The following are the main features of our approach

\begin{itemize}
\item[(1)] The setting allows lattice systems on an infinite cubic lattice $\Ir^s$ of any space dimension $s$, with finite dimensional internal Hilbert space $\Cx^d$. The notion of gauge transformations is established at this kinematical level.
\item[(2)] Electric and magnetic fields are properties of {\it translation systems}, the analogues of infinitesimal space-time translations as described by a connection. We show that translation systems are equivalent (in a natural sense) iff they are connected by a gauge transformation iff they have the same fields.
\item[(3)] At the level of translation systems there is no difference between space and time, so that time just adds one dimension to the spatial lattice. In this way magnetic and general electromagnetic fields are handled in the same framework.
\item[(4)] The cohomology of differential forms has a direct translation to the discrete case \cite{hydon2004variational,mansfield2008difference}, where $p$-forms are functions defined on the set of $p$-dimensional facets of the cubic lattice. Every electromagnetic field arising from a translation system satisfies the discrete analog of the homogeneous Maxwell-equations $dF=0$ and, conversely, any field with this property can be realized in this way. Field zero means gauge equivalence to a trivial translation system.
\item[(5)] Every $p$-form in the continuum setting is mapped to a discrete $p$-form by integration over the appropriate facets. This map commutes with exterior differentiation, but is highly many-to-one. We also construct a continuization map in the opposite direction.
\item[(6)] The analogue of minimal coupling, i.e., putting a given quantum walk into a field is explicitly defined whenever the walk is given as a finite product of subshift and coin operators. Here subshift operators are shifts, that may be conditional on the internal degree of freedom, but do not otherwise affect the internal degree of freedom, and coin operators act at each site separately, possibly in a different way.
\item[(7)] Different decompositions of the same unitary walk operator may lead to different results. This is analogous to the observation in the continuum case that the outcome of minimal coupling depends on ``operator ordering'', i.e., how the Hamiltonian is written as a non-commutative polynomial of momenta. Operator ordering is irrelevant before minimal coupling, since the momenta commute, but makes a real difference afterwards, which cannot be gauged away.
\item[(8)] Constant fields play an important role in practice. As in the continuum case they require that a homogeneous system is described by non-constant potentials. Nevertheless, the translations act as a symmetry up to gauge transformations, and this defines a ``dual'' translation system expressing the symmetry.
\item[(9)] When the field is rational, a regrouping can be used to restore full translation symmetry for a system of supercells.
\end{itemize}

We believe that the close analogies with the continuous case and ordinary electrodynamics sufficiently justify talking of ``electromagnetic fields'' in this context. Of course, these structures survive a continuum limit, by which one hence comes back effortlessly to ordinary electromagnetic fields. Our approach thus has some overlap with work \cite{DebbaschMag,DebbaschLandau,DebbaschGrav} that introduces ``electromagnetic fields'' as a structure turning into proper fields in the continuum limit. Electromagnetic lattice systems are also important for the implementation of quantum simulators \cite{buluta2009quantum,bloch2012quantum,blatt2012quantum}, e.g., for solid state systems. The simulating system may consist of neutral atoms in an optical lattice, and although physical electromagnetic fields are around, and are used for controlling the atoms, these are not the ``simulated'' fields, which have to be implemented in another way. For example, electric fields have been realized by accelerating the lattice \cite{meschede13}. A discussion of the options for magnetic fields in 2D is found in \cite{Sajid}. In any case the justification of calling such a system magnetic is in the realization of the structure we describe, or some version thereof. For the simple systems studied so far a direct Peierls substitution gives the right result, but for more complex walks and cellular automata a systematic approach is called for.

We do not address here the dynamical consequences of electromagnetic fields. One case that is completely understood is that of 1D electric walks \cite{ewalks,locQuasiPer}. Here the long time behaviour and the spectrum depend on the rational/irrational character of the field parameter in the form of its continued fraction expansion. A similarly sensitive dependence is found for 2D magnetic systems, leading to a version of the well-known Hofstadter butterfly \cite{hof76}. General results on propagation, or the analogs of Landau orbits do not seem to exist yet. Fascinating trapping behaviour of the boundary between two regions with different magnetic field, characterized by distinct Chern numbers, is predicted in \cite{Sajid}.

Our paper is organized as follows. We begin by recalling the continuum case (Sect.~\ref{sec:recap}), and describe the kinematical setup for discrete gauge fields in Sect.~\ref{sec:gauge}. In this section only the translations from one space-time point to another are considered, and space and time are treated in exactly the same way. A system Hilbert space (with normalization over space, but not over time) is only introduced in the next Sect.~\ref{sec:NowWalk}. This includes also the equations of motion and the discrete minimal coupling scheme. In the examples section we treat the homogeneous case (Sect.~\ref{sec:homo}), rational fields and regrouping (Sect.~\ref{sec:ratfields}), electric walks in 1D (Sect.~\ref{sec:ewalks}), quasi-periodicity in space and/or time (Sect.~\ref{sec:qperiodic}), and magnetic walks in 2D (Sect.~\ref{sec:butter}).

\section{Minimal coupling principle recalled}\label{sec:recap}

Before delving into the realm of quantum walks, let us briefly review the minimal coupling principle introducing electromagnetic fields to systems continuous in time and space. Consider a $s$-dimensional system with position and momentum operators $Q_k,P_k,\:k=1,\dots s$. Then, the dynamics is implemented by the Schr\"{o}dinger equation $i\partial_t\psi_t=H\psi_t$ where, for simplicity, we take the Hamiltonian $H=h(P_1,\dots,P_s)$ to be a function of momenta alone.
Yet, this equation of motion by itself is not compatible with local gauge transformations $\psi_t\mapsto\psi_t':=e^{i\chi_t(Q)}\psi_t$.

To make up for this omission, gauge potentials are introduced via the \emph{minimal coupling principle}: in the Schr\"{o}dinger equation one substitutes the \emph{canonical} by so-called \emph{kinematical} momentum operators, i.e.
\begin{align}
  P_\mu\mapsto \magtcont\mu&:=P_\mu-A_\mu(t,Q),    \label{kinmom}
\end{align}
where $P_0=i\partial_t$, and $A_0$ as well as the $A_k$ are functions on $\Rl\times\Rl^s$. These substitutions guarantee that gauged solutions to the Schr\"{o}dinger equation are again solutions, but for the Hamiltonian with gauge transformed $A_\mu$, i.e.
\begin{equation}
  \magtcont0\psi_t=\widetilde H\psi_t\quad\Rightarrow\quad \magtcont0'\psi_t'=\widetilde H'\psi_t',
\end{equation}
where $\widetilde H=h(\magtcont1,\dots,\magtcont s)$ and the gauge potentials transform like $A_0\mapsto A_0'=A_0-\partial_0\chi$ and $A_k\mapsto A_k'=A_k+\partial_k\chi$.

Unlike the canonical momentum operators, the $\magtcont\mu$ clearly do not commute anymore. Instead, their commutators 
\begin{equation}\label{comrelcont}
  [\magtcont\mu,\magtcont\nu]=i(\partial_\mu A_\nu-\partial_\nu A_\mu)=iF_{\mu\nu}
\end{equation}
equates to the electromagnetic field-strength tensor encoding electric and magnetic fields.
As one quickly verifies, these fields are invariant under the gauge transformation of the $A_\mu$.

Conversely, for any field $F_{\mu\nu}$ there exists functions $A_\mu,A_\nu$ such that \eqref{comrelcont} holds. This important result follows directly from the Poincar\'e lemma, see Appendix \ref{app:DisCont}, and the homogeneous Maxwell equations
\begin{equation}
  \partial_{[\alpha}F_{\mu\nu]}=0,
\end{equation}
where $[\cdots]$ denotes antisymmetrization of indices. A discrete version of the Poincar\'e Lemma will allow us to prove an analogous statement on the uniqueness of discrete electromagnetic fields below.

\section{Lattice and gauge}\label{sec:gauge}
In this section we set up the basics of gauge theory on a lattice. We follow roughly the well-known continuum theory, and to allow these intuitions to be used more easily we provide a translation table (Table~\ref{tab:concepts}) of basic concepts. The same table applies, when a lattice field is derived from a continuum electromagnetic field, for example in the tight binding approximation (see Sect.~\ref{sec:tight}). But the discrete concepts do not require  such a continuum background. Indeed, a simulated electromagnetic field will rarely be derived in this way, and the implementation of electromagnetic systems and the checking of their properties has to be carried out completely in the discrete setting.

\begin{table}
  \centering
\begin{tabular}{|c|c|c|}
physical concept&continuous GT&discrete GT\\\hline
position space&base manifold& $\Ir^s$\\
&vector bundle  &$\{\HH_x\}_{x\in\Ir^s}$ \\
pure quantum state&bundle section&$\psi\in\bigoplus_x\HH_x$\\
vector potential&connection & translation system\\
field&curvature &\ plaquette operators \\
\end{tabular}
  \caption{Basic concepts of continuum gauge theory (GT) and their lattice analogs}\label{tab:concepts}
\end{table}

\subsection{Translation systems and gauge transformations}\label{sec:tsys}

We begin by describing the kinematical setup of lattice gauge systems, which will be the background for the dynamical evolution by quantum walks. As is well known, gauge theories live in vector bundles, and one can also say that we set up the vector bundle structures needed for discrete electromagnetism. This will be very simple, since the base manifold of the bundle is the lattice $\Ir^s$, so there are no differentiability conditions, and we can work with global bundle charts. The vector space at each point is taken as a Hilbert space $\HH_x$ of the same finite dimension $d$. One can think of each $\HH_x$ as the same fixed Hilbert space $\Cx^d$. But no particular isomorphism is fixed from the outset. Even with all spaces equal, the distinction between different $\HH_x$ is useful as it helps with the book-keeping and indicates where a vector is located.

The basic object we study is the discrete analogue of a connection, and allows us to ``transport'' vectors along lattice directions. When $\alpha\in\{1,\ldots,s\}$ labels the positive lattice directions, we denote by
$\hat\alpha$ the corresponding unit vector. Then a \textbf{translation system} is denoted by $T_1,\ldots,T_s$, where each $T_\alpha$ is a family of unitary operators
\begin{equation}\label{tsys}
   T_\alpha:\HH_x\to\HH_{x+\hat\alpha}, \qquad x\in\Ir^s.
\end{equation}
At this point we could include the parameter $x$ in the notation of $T_\alpha$, e.g., write the above operator as $T_\alpha(x)$, but it turns out to be less cumbersome to keep track of the spaces $\HH_x$, in which the argument of $T_\alpha$ lies, and then pick the appropriate unitary operator.

Just choosing a different basis in each $\HH_x$ changes a translation system only in a trivial way, and we capture this in the following definition. By a \textbf{localized operator} we mean a collection of operators $A(x)$ with $A(x)$ acting in $\HH_x$. For a localized \emph{unitary} operator $U$, every $U(x)$ is unitary. These are the local basis changes, possibly depending on $x$. For the unitary equivalence in the following definition it is convenient to allow the spaces $\HH_x$ resp.\ $\HH'_x$ to be different as well, so $U$ becomes a family of unitary operators $U(x):\HH_x\to\HH_x'$.

\begin{defi}\label{def:isotranslation}
Two translation systems $T,T'$ on respective families  of Hilbert spaces $\{\HH_x\}_{x\in\Ir^s}$ and $\{\HH'_x\}_{x\in\Ir^s}$ are called \textbf{equivalent} if there is a localized unitary operator $U$ such that $UT_\alpha U^*=T'_\alpha$.
\end{defi}

Note that in this operator product each of the factors is really a family of unitary operators, acting in an $x$-dependent way. Thus
\begin{equation}\label{UTU}
  (UT_\alpha U^*)\phi_x'=U(x+\hat\alpha)T_\alpha U(x)^*\phi_x'.
\end{equation}

A typical way to detect non-equivalence is to transport a vector around a closed path, which may lead to a different vector. To build such paths we also need steps in the $-\hat\alpha$ direction. In the shorthand notation for paths, where we just keep track of the directions, we label these with $\{-1,\ldots,-s\}$. The corresponding unitary step operator is then $T_{-\alpha}=T_\alpha^*$.
A path on the lattice is defined by a sequence $\gamma=(\gamma_1,\ldots\gamma_\ell)$, with $\gamma_i\in\{\pm1,\ldots,\pm s\}$. Naturally we set
\begin{equation}
  T_\gamma=T_{\gamma_{\ell}}\cdots T_{\gamma_1}.
\end{equation}
Closed paths, or  ``loops'', are those whose net transport $\hat\gamma=\sum_i\hat{\gamma_i}$ vanishes. For such operators $T_\gamma:\HH_x\to\HH_x$ for every $x$, i.e., $T_\gamma$ is a localized unitary operator.
The loop operators define the \textbf{holonomy group} at $x$ as
\begin{equation}\label{holo}
  {\hol}_x=\{T_\gamma(x)| \gamma\ \text{ a loop}\}.
\end{equation}
We call a translation system \textbf{flat} if the holonomy group consists everywhere only of the identity. That is, the transport of a vector $\phi_x\in\HH_x$ to $\HH_y$ by an operator $T_\gamma$ with $x+\hat\gamma=y$ is independent of the path $\gamma$. Clearly, only a flat translation systems allows a meaningful interpretation of ``$\phi_x\in\HH_x$ and $\psi_y\in\HH_y$ are equal''.
We can use this to rewrite any given translation system in a simplified but equivalent form.

\begin{lem}\label{lem:flat}
	For every translation system $T$ we can find a flat system $\Tf$ such that
		\begin{equation}\label{Uax}
		T_\alpha\phi_x=\Tf_\alpha U_\alpha(x)\phi_x
		\end{equation}
		with a unitary $U_\alpha(x)\in\hol_x$. Moreover, there is an equivalent translation system $T'$, where ${\hol'_x}$ is the same for all $x$ with respect to $\Tf$.
\end{lem}

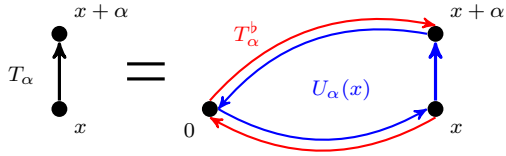
\begin{figure}
%

\tikzset{
	>=stealth',
	right graphic/.style={
		xshift=4.5cm
	},
	left graphic/.style={
	xshift=-5cm
}
}

\begin{tikzpicture}[font=\footnotesize]

\def\plusx{4}
\def\plusy{1.5}
\def\plusl{.4}


\node[left graphic,circle,fill=black,inner sep=0pt,minimum size=6pt, label=south east:{$x$}] (px) at (3,1) {};

\node[left graphic,circle,fill=black,inner sep=0pt,minimum size=6pt, label=north east:{$x+\alpha$}] (pxa) at (3,2) {};

\path[left graphic,->,very thick] (px) edge (pxa);

\draw[very thick] (-1.1,\plusy+\plusl/4) -- (-.6,\plusy+\plusl/4) ;
\draw[very thick] (-1.1,\plusy-\plusl/4) -- (-.6,\plusy-\plusl/4) ;

\node[left graphic] (label) at (2.5,1.45) {$T_\alpha$};


\node[circle,fill=black,inner sep=0pt,minimum size=6pt, label=south west:{$0$}] (p0) at (0,1) {};

\node[circle,fill=black,inner sep=0pt,minimum size=6pt, label=south east:{$x$}] (px) at (3,1) {};

\node[circle,fill=black,inner sep=0pt,minimum size=6pt, label=north east:{$x+\alpha$}] (pxa) at (3,2) {};

\path[->,blue,thick] (p0.east) edge [bend right] (px.west);
\path[->,blue,very thick] (px) edge (pxa);
\path[->,blue,thick] (pxa.west) edge [bend right] (p0.east);
\path[<-,red,thick] (p0.south) edge [bend right] (px.south);
\path[<-,red, thick] (pxa.north) edge [bend right] (p0.north);

\node (label1) at (1.75,1.25) {\textcolor{blue}{$U_\alpha(x)$}};
\node (label2) at (.5,2) {\textcolor{red}{$\Tf_\alpha$}};

\end{tikzpicture}

	\caption{\label{fig:flatloop} Construction of the flat translation system $\Tf$ in the proof of \ref{lem:flat}, which allows to express a given translation system $T$ by multiplication of an element of the holonomy group of $T$ (blue) followed by $\Tf$ (red), see \eqref{Uax}.}
\end{figure}

\begin{proof}
	The key element to chose here is, for every $x$, a path $\gamma(x)$ from a reference point, say $0$, to $x$, such that $T_{\gamma(x)}:\HH_0\to\HH_x$. The flat translation system may now be defined by
	\begin{equation}\label{tflateq}
	\Tf_\alpha \phi_x= T_{\gamma(x+\hat\alpha)}T_{\gamma(x)}^*\phi_x.
	\end{equation}
	This is flat, because for any path $\Tf_\gamma$, during which some sequence of lattice points is traversed, it corresponds to going back to $0$ in every step, and then forward to the next point. These jumps cancel from step to step.
	With this we can write
	\begin{align}
	T_\alpha\phi_x=T_{\gamma(x+\alpha)}T_{\gamma(x)}^*T_{\gamma(x)}T_{\gamma(x+\alpha)}^*T_{\alpha}\phi_x,
	\end{align}
	which evaluates to \eqref{Uax}, with $U_\alpha(x)=T_{\gamma(x)}T_{\gamma(x+\alpha)}^*T_{\alpha}\in\hol_x$
	 (compare Fig.~\ref{fig:flatloop}).
	To prove the equivalence of all $\hol_x$, we use the standard paths $\gamma(x)$ to identity the spaces $\HH_x$ with each other. The unitary equivalence operator will be $V(x):\HH_x\to\HH_x'=\HH_0$, defined as
	\begin{equation}
		V(x)\phi_x=T_{\gamma(x)}^*\phi_x.
	\end{equation}
	Given now an element of $\hol_x$, i.e.\ a unitary $T_\sigma$ with a closed loop $\sigma$, starting at $x$, it transforms to
	\begin{equation}
		T_\sigma'\phi_x'=VT_\sigma'V^*\phi_x'=T_{\gamma(x)}^*T_{\sigma}T_{\gamma(x)},
	\end{equation}
	which is describes a loop, starting at $x=0$. Hence, $\hol'_x=\hol_0$ for all $x$.
\end{proof}

	Note that after the identification $\HH_x'\equiv\HH_0$ the flat translation system \eqref{tflateq} not only has trivial $\hol_x$, but also acts as the identity for all $\phi_x'$
\begin{equation}
(\Tf_\alpha)'\phi'_x=\phi'_{x+\alpha}.
\end{equation}

This lemma underlines the importance of the holonomy group. Typically one actually constrains this group to be a subgroup of a certain group $\ggroup\subset U(d)$, the ``gauge group'', and fixes also the form \eqref{Uax} with unitaries from $\ggroup$. This defines a family of translation systems, for many of which $\hol_x$ will be dense in $\ggroup$. In this sense a single translation system and its holonomy group may determine the whole gauge family. The resulting restricted kind of translation systems is described in the next section, before we further specialize to electromagnetic fields, for which $\hol_x$ consists of phases only.

\subsection{Gauge transformations}\label{sec:gaugeTrafo}
Following the structure found in Lemma~\ref{lem:flat} we now fix a flat translation system $\Tf$, and hence an a priori identification of the spaces $\HH_x\equiv\Cx^d$, which then only differ by their location index. Moreover, we fix a \textbf{gauge group} $\ggroup$, as a subgroup of the unitary group of $\Cx^d$. Then a \textbf{$\ggroup$-translation system} is one of the form
\begin{equation}\label{UaxG}
   T_\alpha\phi_x=\Tf_\alpha U_\alpha(x)\phi_x,\quad \forall x,\alpha:\: U_\alpha(x)\in\ggroup.
\end{equation}
A \textbf{gauge transformation} is a localized unitary $V$ with $V(x)\in\ggroup$, and two translation systems are called \textbf{gauge equivalent} if the operator $U$ in Def~\ref{def:isotranslation} can be chosen as a gauge transformation.

The structure of a translation system is now encoded in the operators $U_\alpha(x)$. There is no constraint on these operators. By multiplying steps we get, for every path $\gamma$, operators $U_\gamma(x)\in\ggroup$  so that
\begin{eqnarray}
  T_\gamma\phi_x &=&\Tf_\gamma U_\gamma(x)\phi_x,\quad \phi_x\in\HH_x \\
  U_{\gamma_1\gamma_2}(x) &=& U_{\gamma_1}(x+\hat\gamma_2)U_{\gamma_2}(x)
\end{eqnarray}
In particular, $\hol_x\subset\ggroup$ for all $x$. This is always a countable, hence proper subgroup of $U(1)$, or even a discrete subgroup (see Sect.~\ref{sec:ratfields}).

Lemma~\ref{lem:flat} suggests a choice of gauge by choosing a suitable set of standard paths $\gamma(x)$ connecting $0$ with any given point $x$. A frequently useful choice is the \textbf{path-ordered gauge} where one first does  all steps in the positive or negative $1$-direction, then all along $2$ and so on.
Thus for $x=(x_1,\ldots,x_s)$ the standard path $\gamma(x)$ leads to
\begin{equation}\label{pathorder}
  T_{\gamma(x)}:=T_s^{x_s}\cdots T_2^{x_2}T_1^{x_1}.
\end{equation}
With the gauge transformation as in the proof of Lemma~\ref{lem:flat} translations along standard paths are used for the identification of neighbouring $\HH_x$, so no additional unitaries are picked up. In particular, the translations along $T_1$ are flat. For $T_2$ we get unitaries $U_2(x)$ depending only on $x_1$, and so on for the further directions.

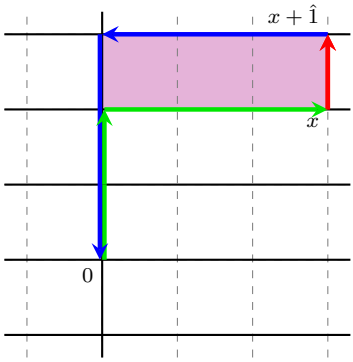
\begin{figure}


\begin{tikzpicture}[font=\footnotesize]
    \def\mx{2.3}
    \def\my{2.3}
    \def\dif{0.03}
    \definecolor{kardinal}{RGB}{168,0,128}

    \fill[kardinal!30!white] (-1,1) -- (2,1) -- (2,2) -- (-1,2) -- cycle;
    \draw[gray,thin,dashed] (-\mx,-\my) grid (\mx,\my);

    \draw[black,thick] (-\mx,-2) -- (\mx,-2) (-\mx,-1) -- (\mx,-1) (-\mx,0) -- (\mx,0) (-\mx,1) -- (\mx,1) (-\mx,2) -- (\mx,2);
    \draw[black,thick] (-1,-\my) -- (-1,\my);

    \draw[green!90!black,line width=1.7] (-1+\dif,-1) -- (-1+\dif,0);

    \draw[red,line width=1.7,>=stealth,->] (2,1) -- (2,2);

    \draw[green!90!black,line width=1.7,>=stealth,->] (-1+\dif,1) -- (2,1);

    \draw[blue,line width=1.7,>=stealth,<-] (-1-\dif,-1) -- (-1-\dif,2);
    \draw[blue,line width=1.7,>=stealth,<-] (-1,2) -- (2,2);

    \draw[red,line width=1.7,>=stealth,->] (2,1) -- (2,2);

    \draw[green!90!black,line width=1.7,>=stealth,->] (-1+\dif,0) -- (-1+\dif,1);

    \draw[below left] (-1,-1) node {$0$};
    \draw[below left] (2,1) node {$x$};
    \draw[above left] (2,2) node {$x+\hat1$};
\end{tikzpicture}
  \caption{\label{fig:pathorder}The path-ordered gauge is a particular example for a maximal tree gauge \cite{creutz1977gauge}, for which no phases are picked up along the black links. In this gauge, the phases picked up along the dashed lines corresponds to the sum of the plaquette phases enclosed by the loop $0\to x\to x+\hat\alpha\to0$ where $0\to x$ and $x+\alpha\to0$ are standard paths.}
\end{figure}

Applying a gauge transformation $V$ the translation system \eqref{UaxG} is transformed to $T'_\alpha=VT_\alpha V^*$, characterized by the operators
\begin{equation}\label{Ugauged}
  U'_\alpha(x)=V(x+\hat\alpha)U_\alpha(x)V^*(x).
\end{equation}
In particular, for a closed path we get
\begin{equation}\label{Ugagauged}
  U'_\gamma(x)=V(x)U_\gamma(x)V^*(x).
\end{equation}
In an abelian gauge theory, i.e., with $\ggroup$ abelian, $V(x)$ can be commuted through, so the $U_\gamma(x)$ do not change under gauge transformations.
Even in non-abelian gauge theories we can get a gauge invariant quantity out of this, namely $\tr U_\gamma(x)$. Note that in either case the invariance covers also the choice of the initial point: For all points on a closed loop on the lattice is the same (when $\gamma$, as the list of directions is also shifted accordingly).

Non-trivial transport around a closed path is the hallmark of curvature. In an abelian gauge theory this quantity simply adds up over surfaces. If we have a large surface $\mathcal S$ divided into two pieces $\mathcal S_1$, $\mathcal S_2$ by a path, we can turn the loop around $S$ into the sum of the loops around $\mathcal S_1$ and $\mathcal S_2$. The dividing path is traversed twice, but in opposite direction, and since all $U_\alpha$ commute, these contributions cancel. General loops can therefore be reduced to elementary ones. In the continuum they are taken to be infinitesimal, given the curvature $2$-form. In the discrete case the smallest we can do is individual plaquettes
\begin{equation}\label{eq:plaquetteoperator}
  P_{\alpha\beta}=T_\alpha^*T_\beta^*T_\alpha T_\beta.
\end{equation}
Note that these operators still depend on the starting point on the lattice. Explicitly, $P_{\alpha\beta}\phi_x=P_{\alpha\beta}(x)\phi_x$, with
\begin{equation}\label{plaqx}
  P_{\alpha\beta}(x)= U_\alpha(x)^*U_\beta(x+\hat\alpha)^*U_\alpha(x+\hat\beta)U_\beta(x),
\end{equation}
where we have used the identity $U_{-\alpha}(x)=U_\alpha(x-\hat\alpha)^*$.

In general, equality of the plaquette traces is not sufficient for $\ggroup$-translation systems to be gauge equivalent. However, this will be true in the case of main interest for us, to which we now turn.

\subsection{U(1)-gauge theory}\label{sec:U1}
We now specialize to the case relevant for electromagnetism, that is, the gauge group $\ggroup=U(1)$, so that $U_\alpha(x)$ for all $x,\alpha$ corresponds to multiplication by suitable phases. The structure of $U(1)$-translation systems is completely determined by these phases, and hence independent of $d$. Likewise, the plaquette operators are just phases, so that  $P_{\alpha\beta}(x)\in U(1)$. By definition these are multiplicatively antisymmetric, i.e,
$P_{\alpha\beta}(x)P_{\beta\alpha}(x)=1$ for all $x$.
The main reason why we can say much more in this case is that there is a cohomology theory for these phase systems:

\begin{thm}\label{thm:uniqueness}\strut\hfil
\begin{itemize}
\item[(1)] Two $U(1)$-translation systems $T,T'$ are gauge equivalent iff their plaquette phases are everywhere equal, i.e.,  $P_{\alpha\beta}(x)=P'_{\alpha\beta}(x)$ for all $x$.
\item[(2)] A system of antisymmetric plaquette phases can arise from some $U(1)$-translation systems if and only iff
    \begin{equation}\label{eq:disc_maxwell}
      \prod_{{\rm cyc}(\alpha\beta\mu)} P_{\alpha\beta}(x+\hat\mu)P_{\beta\alpha}(x) =1,
    \end{equation}
where the product is over all cyclic permutations of the indices $(\alpha,\beta,\mu)$.
\end{itemize}
\end{thm}

In the continuum the first equation is equivalent to the statement that two vector potentials give the same field iff they differ by a gradient, and the second corresponds to the homogeneous Maxwell equations. To see these analogies, it helps to write the abelian group additively, i.e., to parameterize phases by the phase angle in $\Rl/(2\pi\Ir)$. That is, for the localized unitaries implementing gauge transformations we write $V(x)=\exp (i\chi(x))$ with $\chi:\Ir^s\to\Rl/(2\pi\Ir)$. Similarly, translation systems are encoded by $U_\alpha(x)=\exp (iA_\alpha(x))$ for a family of functions $A_\alpha:\Ir^s\to\Rl/(2\pi\Ir)$, and plaquette phases are given as $P_{\alpha\beta}(x)=\exp(iF_{\alpha\beta}(x))$. Denoting by $d_\alpha$ the discrete derivative in direction $\alpha$, i.e.,
\begin{equation}\label{latderivative}
 (d_\alpha f)(x)=f(x+\hat\alpha)-f(x),
\end{equation}
\eqref{plaqx} translates to
\begin{equation}\label{plaqxf}
  F_{\alpha\beta}(x)=d_\alpha A_\beta(x)-d_\beta A_\alpha(x),
\end{equation}
the discrete analogue of the electromagnetic field-strength tensor \eqref{comrelcont}.

\begin{proof}
  To prove Theorem \ref{thm:uniqueness} we set up a discrete differential calculus along the lines of \cite{hydon2004variational,mansfield2008difference}, see Appendix \ref{app:DisCont} for details. In this parlance of discrete differential calculus the $A_\alpha:\Ir^s\to\Rl/(2\pi\Ir)$ implementing a translation system define a discrete $1$-form
  \begin{equation}
    A=\sum_\alpha A_\alpha\dd x^\alpha.
  \end{equation}
  Similarly, $F_{\alpha\beta}:\Ir^s\to\Rl/(2\pi\Ir)$ defines a discrete $2$-form
  \begin{equation}
    F=\sum_{\alpha<\beta} F_{\alpha\beta}\dd x^\alpha\wedge\dd x^\beta.
  \end{equation}
  A discrete exterior derivative is defined in \eqref{eq:ddisc}, and we call discrete forms which are given as $\omega=d\eta$ \emph{exact} and such for which $d\omega=0$ \emph{closed}. Clearly, $d$ satisfies $d^2=0$, i.e. every exact form is closed.

  Then, the first part of the theorem follows from
  \begin{equation}
    F=dA,
  \end{equation}
  which is equivalent to \eqref{plaqxf}.

  That every $U(1)$-translation system satisfies \eqref{eq:disc_maxwell}, follows from direct calculation, and is equivalent to
  \begin{equation}
    dF=0.
  \end{equation}
  The converse statement, that every set of antisymmetric plaquette phases satisfying \eqref{eq:disc_maxwell} arises from a $U(1)$-translation system is a special instance of the general statement ``every closed form is exact''. This is the content of the discrete Poincar\'e lemma which we discuss in Appendix \ref{app:DisCont}.
\end{proof}

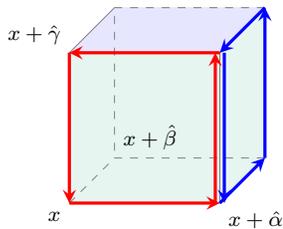
\begin{figure}[t]
%
%
%

\begin{tikzpicture}[font=\footnotesize,scale=2]
    \tikzstyle{arvt} = [>=stealth,very thick]
    \tikzstyle{art} = [>=stealth,thick]
    \tikzstyle{gr} = [gray,very thin]
    \tikzstyle{grdash} = [gray,very thin,dashed]

    \coordinate (000) at (0,0);
    \coordinate (100) at (1,0);
    \coordinate (010) at (.3,.3);
    \coordinate (110) at (1.3,.3);
    \coordinate (001) at (0,1);
    \coordinate (101) at (1,1);
    \coordinate (011) at (.3,1.3);
    \coordinate (111) at (1.3,1.3);

    \fill[blue!10!white] (000) -- (100) -- (110) -- (010) -- cycle;
    \fill[blue!10!white] (000) -- (010) -- (011) -- (001) -- cycle;
    \fill[blue!10!white] (010) -- (110) -- (111) -- (011) -- cycle;

    \fill[green!10!white,opacity=.6] (000) -- (100) -- (110) -- (111) -- (101) -- (001) -- cycle;

    \draw[gr] (000) -- (100) -- (101) -- (001) -- (000);
    \draw[grdash] (000) -- (010) -- (110) -- (111) -- (011) -- (010);
    \draw[gr] (001) -- (011) (100) -- (110) (101) -- (111)  ;

    \draw[arvt,->,red] (000) -- (100);
    \draw[arvt,->,red] (.97,0) -- (.97,1);
    \draw[arvt,->,red] (101) -- (001);
    \draw[arvt,->,red] (001) -- (000);

    \draw[arvt,->,blue] (100) -- (110);
    \draw[arvt,->,blue] (110) -- (111);
    \draw[arvt,->,blue] (111) -- (101);
    \draw[arvt,<-,blue] (1.03,0) -- (1.03,1);

    \node[below left] at (000) {$x$};
    \node[below right] at (100) {$x+\hat\alpha$};
    \node[above right] at (010) {$x+\hat\beta$};
    \node[above left] at (001) {$x+\hat\gamma$};

\end{tikzpicture}
  \caption{\label{fig:plaquette3D}A visualization of the discrete Maxwell equations \eqref{eq:disc_maxwell}. Each of the links of the boundary of this cube appear twice, once in positive and once in negative direction, and the corresponding contributions cancel.}
\end{figure}

\subsection{Continuous to discrete}\label{sec:tight}
The standard application of lattice gauge theory in solid state systems is to describe electrons moving under the combined influence of the periodic potential provided by some positively charged ions of a crystal lattice, and an external electromagnetic field. In an approximation of non-interacting electrons the system without external fields is essentially a one-electron Hamiltonian system in a periodic potential. It can be diagonalized jointly with the lattice translations and thus analyzed as a function of quasi-momentum (the Brillouin zone). Since the kinetic energy is unbounded, this gives infinitely many bands, and restricting to the lowest bands (or selecting a small set of electron wave functions per atom) one gets the tight binding model, in which the electrons are described by a discrete lattice position plus some internal degree of freedom.

Let us now turn on an external electromagnetic field, which is given in terms of $A_\mu(x)$, $\mu=0,\ldots,s$, $x\in\Rl^{s+1}$. How does this lead to a ``lattice gauge field'' in the tight binding approximation? This is easy enough for electric potentials in a suitable gauge, which we can apply as multiplication operators just as in the continuous case. The only thing that changes is that we now have to evaluate $A_0$ only at the lattice points. In other words, a scalar potential $A_0$ is discretized as multiplication by $\exp(iA_0(x))$ on $\ell^2(\Ir^s)$.

However, for vector potentials this method is bad, because it does not reflect their meaning as a connection, i.e., as a structure that encodes a notion of parallel transport. When $\partial_\alpha-iA_\alpha$ is the generator of translations in the continuum, it is natural to take its lattice analog as the phase acquired in the continuum theory by transporting a vector from $x$ to $x+\hat\alpha$. That is
\begin{equation}\label{Peierls}
  U_\alpha(x)=\exp i\int_0^1 dt\ A_\alpha(x+t\hat\alpha),
\end{equation}
and is called the \textbf{Peierls phase} \cite{Peierls}. Clearly, this differs from $\exp( iA_\alpha(x))$, the naive expectation from the case of electric fields. When $A_\alpha=\partial_\alpha\chi$, \eqref{Peierls} amounts to $U_\alpha(x)=\exp \bigl(i(\chi(x+\hat\alpha)-\chi(x))\bigr)$, i.e., the lattice gradient. So this kind of discretization is compatible with the differential calculus.

For $2$-forms, the interpretation of the field as curvature $F_{\alpha\beta}$ suggests to take integrals over plaquettes, and again this makes the definition compatible with differentiation. In general, a $p$-form
is an antisymmetric tensor with $p$ indices. Its discrete version will be an integral over a $p$-dimensional face of the lattice, and one might think of it as a quantity associated with a plaquette: If $\alpha,\beta,\ldots$ are the indices of the electromagnetic field-strength tensor, the plaquette is labelled by a lattice point, from which it is spanned by the vectors in the positive $\hat\alpha,\hat\beta,\hat\gamma,\ldots$ directions. Note that this matches the definition \eqref{latderivative} of the lattice derivatives $d_\alpha$ as unilateral differences in the positive direction. Appendix \ref{app:DisCont} shows that the discretization thus intertwines exterior differentiation with the lattice exterior derivative. There is even a similar map in the opposite direction. However, this cannot be an inverse, because the discretization map sketched here is clearly many-to one. For example, the discretization of a vector potential which is nonzero only on the interior of plaquettes is simply zero.

\section{Discrete Minimal coupling}\label{sec:NowWalk}
\subsection{The Hilbert space}

So far we have described only the kinematical part of the theory: a bundle-like structure of Hilbert spaces attached to points $x\in\Ir^s$ with a system of translations. There was no dynamics and no time coordinate.

We start by singling out a time coordinate. Thus our lattice becomes $\Ir\times\Ir^s=\Ir^{s+1}$, with the first coordinate henceforth playing the role of time. The point here is that at the level of translation systems there is no difference between temporal and spatial translations, and therefore we automatically have the typical combination of electric field components $F_{0k}$ and magnetic components $F_{k\ell}$, where $k,\ell=1,\ldots,s$ are the spatial indices. We will write the points of $\Ir\times\Ir^s$ as pairs $(t,x)$, and $\HH_{t,x}$ for the local Hilbert space at the event $(t,x)$.

The difference between time and space comes in when we describe the Hilbert space of the system. This will now be a family of Hilbert spaces indexed by time, namely
\begin{equation}\label{timeslice}
  \HH(t)=\bigoplus_{x\in\Ir^s}\HH_{t,x}.
\end{equation}
This direct sum is to be read as a Hilbert space direct sum with the norm $\norm{\psi(t)}^2=\sum_x\norm{\psi(t,x)}^2$, and associated scalar product. Note that $\psi(t,x)$ is the component of $\psi(t)$ in the direct summand $\HH_{t,x}$. There is no normalization condition involving a sum over time. This definition makes sense in the abstract setting, but it can be simplified to
\begin{equation}\label{HHtcoo}
  \HH(t)=\ell^2(\Ir^s)\otimes\Cx^d
\end{equation}
after identifying the local Hilbert spaces by a flat translation system.

Let us consider some operators in these Hilbert spaces.
Every localized operator $A$ (see Sect.~\ref{sec:tsys}) simply acts on $\HH(t)$ as $\bigoplus_xA_{t,x}$, i.e., for $\psi(t)\in\HH(t)$ we have $(A\psi)(t,x)=A_{t,x}\psi(t,x)$. This means that localized operators become a time-dependent family of operators $A(t)$ acting in $\HH(t)$. In particular, this goes for gauge transformations.

The spatial translations of a translation system become a family of unitary operators $\Ts_k\in\BB(\HH(t))$, acting as
\begin{equation}\label{unitarytrans}
  (\Ts_k\psi)(t,x)=T_k\Bigl(\psi(t,x-\hat k)\Bigr).
\end{equation}
Here the arguments are dictated by the rule that $\psi(t,x)\in\HH_{t,x}$ also holds for $\Ts_k\psi$.

A \textbf{constant operator} $A$ is a localized operator commuting with all translations. Since for a $U(1)$-theory all localized operators commute with gauge transformations, a constant operator remains constant for any translation system differing by a gauge transformation as in Sect.~\ref{sec:gaugeTrafo}. In the factorization \eqref{HHtcoo} the constant operators are then simply the ones of the form $\idty\otimes A$ with $A$ an operator on $\Cx^d$

\subsection{Dynamics: Walks}

A pure quantum state is now given by a family of vectors $\psi(t)\in\HH(t)$, which we interpret as the state ``at time $t$''. Knowing the state at one time allows us to determine it for all times. This is expressed by a \textbf{dynamical constraint}, or \textbf{equation of motion}. This  connection between $\psi(t)$ and $\psi(t+1)$ will involve $T_0$, the timelike member of the translation system, which maps $\HH(t)$ to $\HH(t+1)$. So one possible dynamical constraint would be $\psi(t+1)=T_0\psi(t)$, which from the point of view of the given translation system just means that $\psi$ is ``constant'' in time. In a more general setting, the dynamical constraint is given by a family of unitary operators $W(t)$ on $\HH(t)$ such that
\begin{equation}\label{dynamics}
  \psi(t+1)=T_0\ W(t)\ \psi(t).
\end{equation}
We have seen that, e.g., in a path ordered gauge the first coordinate can always be represented by a flat translation. Choosing such a \emph{temporal gauge} allows us to just identify the Hilbert spaces $\HH(t)$, and drop the map $T_0$ from the equation. In this way we go back to the more common description in a fixed Hilbert space with a possibly time-dependent unitary step operator $W(t)$. The reason for choosing the form \eqref{dynamics} is that it makes clear how to include gauge transformations which affect the temporal component. What does not work is to take the operator $W$ itself as the time translation. This is like confusing $i\partial_t$ with the Hamiltonian. In any case, since $W$ typically spreads the wave function to many sites, this would not fit the description of translation systems, on which the unification of magnetic and electromagnetic case is based.

\subsection{Walks, shifts, and coins}
We typically require that $W$ has finite \textbf{jump length} $L$, which means that $W(t)\HH_{t,x}\subset\bigoplus_{y;\,|y-x|<L}\HH_{t,y}$. Here we briefly describe some standard ways for writing down a walk.

Let us first consider the case without external fields, using the flat translation system only, and just one fixed $t$. We can write $W\equiv W(t)$ as
\begin{equation}\label{Wkernel}
  (W\psi)(x)=\sum_yW(x,y)\psi(y),
\end{equation}
where the kernel $W(x,y):\HH_y\to\HH_x$ is operator valued. When $\HH_x\equiv\Cx^d$, each $W(x,y)$ just a $d\times d$-matrix. These matrices by themselves are typically not unitary because they contain only one part of the jump amplitudes. Instead unitarity, i.e., preservation of probability, is expressed by a sum involving all jumps.
Rather than the jump origin (as in \eqref{Wkernel}), we can take the translation $z=(x-y)$ to index the sum, emphasizing the translation part of such a map, by bringing in the unitary shift operators $\Tsf_\alpha$ (see \eqref{unitarytrans}):
\begin{equation}\label{walk0}
  (W\psi)(x)=\sum_z W_z(x)(\Tsff z\psi)(x)
\end{equation}
so that $W_z$ becomes a localized operator, and the sum is only over $|z|<L$. Here $\Tsff z$ with $z=(z_1,\ldots,z_s)$ is a shorthand for $(\Tsf_1)^{z_1}\cdots(\Tsf_s)^{z_s}$, the shift along the lattice vector $z$. Formula \eqref{walk0} is a matrix-valued expression in the following sense: We can think of $\ell^2(\Ir^s)\otimes\Cx^d$ either as $\ell^2(\Ir^s;\Cx^d)$, that is vectors which are functions on $\Ir^s$ with ``spinor'' values in $\Cx^d$, or else as $d$-component vectors, whose entries are in $\ell^2(\Ir^s)$. In the latter view we can think of a walk as a $\d\times d$-block matrix operator whose entries are operators on $\ell^2(\Ir^s)$. Then in the above formula each $W_z(x)$ is a $d\times d$-matrix, so the same formula holds matrix element by matrix element. Consider, for example, the translation invariant walk, which is the basis of the example in Sect.~\ref{sec:butter}.
Its matrix can be written as
\begin{eqnarray}\label{tiwalk}
   W&=&\frac12\begin{pmatrix}\Tsf_2&0\\0&\Tsff*_2\end{pmatrix}\begin{pmatrix}1&1\\1&-1\end{pmatrix}\begin{pmatrix}\Tsf_1&0\\0&\Tsff*_1\end{pmatrix}\begin{pmatrix}1&1\\1&-1\end{pmatrix} \nonumber\\
    &=& \frac12\begin{pmatrix}\Tsf_2(\Tsf_1+\Tsff*_1)&\Tsf_2(\Tsf_1-\Tsff*_1)\\\Tsff*_2(\Tsf_1-\Tsff*_1)&\Tsff*_2(\Tsf_1+\Tsff*_1)\end{pmatrix}.
\end{eqnarray}
Here every matrix entry is a polynomial in the shift operators. It has constant coefficients, which expresses translation invariance. For more general space dependent coins there would also be $x$-dependent coefficients.

The first form in \eqref{tiwalk} is often used in the (theoretical or experimental) construction of walks, and is called a shift-coin decomposition. It is a product of two kinds of operations: On the one hand, the  ``coin'' operations are just the localized unitary operators. The term ``coin'' arose in the Quantum Information community, where a spatially constant localized operation was introduced as the analogue of flipping a coin to decide in which direction the system would move by the next step. The steps are implemented by shifts, but it is crucial to allow the shifts to depend on the internal (``coin''-) state of the system. Therefore we must also allow \textbf{subshifts} of the form
\begin{equation}\label{subshift}
 \Ts^P_k= P\Ts_k+(\idty- P)=\begin{pmatrix}
                              \Ts_k & 0 & 0 & 0 \\
                              0 & 1 & 0 & 0 \\
                              0 & 0 & \ddots & 0 \\
                              0 & 0 & 0 & 1
                            \end{pmatrix},
\end{equation}
where $P$ is a constant projection, which in the second equality has been taken as $\kettbra1$.
This special case suffices, because we allow products, and by conjugation with a constant coin we can include also shifts with $P=\kettbra\phi$, or like the first factor in \eqref{tiwalk}.

A walk will be called \textbf{decomposable}, if it can be written as a finite product
\begin{equation}\label{walkDec}
  W=C_0S_1C_1S_2\cdots S_nC_n
\end{equation}
of subshifts $S_i=\Ts^{P_i}_{k_i}$ and coin operations $C_i$. It is not known, whether this comprises all walks with finite jump length, but certainly covers all examples in the literature.

\subsection{Walks in a minimally coupled external field}\label{sec:mincoup}
Let us begin with some walk in zero field. How can we consider ``the same'' walk in an external field? Roughly speaking, minimal coupling is done by replacing the flat translation system in \eqref{walk0} by a general one, expressing the field in question. That sounds easy enough, but the powers in \eqref{walk0} are now ambiguous, and would depend on the choice of a path from $0$ to $z$. Hence the basic minimal coupling scheme has to be augmented by a scheme of choosing a path for every power appearing in \eqref{walk0}, possibly a different one in different matrix elements of $W$. Changing the path ordering in any one of these places is likely to ruin unitarity. Therefore, a rather subtly connected set of choices is demanded.

There is one scheme, however, which immediately takes care of all these choices: One makes the substitution not in the form \eqref{walk0} but in a fixed decomposition \eqref{walkDec} of the walk into subshifts and coins. The coins are not affected, and for the subshifts it amounts to replacing the upper left diagonal block matrix $\Ts_k$ in \eqref{subshift} by another unitary operator. Clearly, this automatically preserves unitarity. Multiplying out the product \eqref{walkDec} one gets back to the form \eqref{walk0}, but now every term comes with a definite operator ordering (compare the two expressions in \eqref{tiwalk}). It is clear that no substitution scheme on the basis of just \eqref{walk0} is likely to handle all these choices coherently. The following definition summarizes the main message of this paragraph.

\begin{defi}\label{def:mincoup}
Consider a quantum dynamical system determined by the equation of motion \eqref{dynamics}, with a walk operator $W$ decomposed into a product \eqref{walkDec} of localized unitary operators and subshifts with respect to some flat translation system $\Tf$. Then the {\bf minimal coupling} of the system to an external field described by a translation system $T$ corresponds to replacing throughout $\Tf_\alpha$ with $T_\alpha$ ($\alpha=0,\ldots,s$) in the equation of motion, and  every subshift.
\end{defi}

The field that is turned on in this way has \textbf{electric components} $P_{0\alpha}$ and \textbf{magnetic components} $P_{\alpha\beta}$, for $\alpha,\beta=1,\ldots s$. Since the $T_\alpha$ are determined by these data up to a choice of gauge, so is the walk, see Sect.~\ref{sec:gaugeEqW} for details.

As examples, we briefly discuss two special cases: In the \textbf{purely electric} setting the the spatial plaquette operators $P_{k\ell}$ are all equal to the identity. This allows us to choose a gauge in which $T_k=\Tf_k$ for $k=1,\ldots,s$. In this gauge $T_0=\Tf_0 U_0(t,x)$, and the electric field in direction $k$ is determined by $P_{0k}=U_0(t,x)^*U_0(t,x+\hat k)$. Note that in the continuous setting the inhomogeneous Maxwell equations imply that this electric field is time-independent. Since we do not have a discrete equivalent of the inhomogeneous Maxwell equations at our disposal we cannot conclude an analogous statement.

In the \textbf{purely magnetic} case where $P_{0k}=\idty$, the homogeneous Maxwell equations imply that the magnetic field is time-independent, which in the continuous setting follows from Faraday's law of induction. More than that, in temporal gauge already the $U_k$ must be time independent by $P_{0\ell}=\idty$.

\subsection{Gauge equivalent walks}\label{sec:gaugeEqW}
We claimed that gauge equivalent minimal coupling substitutions give equivalent walks. One could express this by saying that the equation of motion is gauge invariant in a natural sense. Let us state this a bit more formally.

\begin{lem}\label{lem:gaugeeq} Let $W$ be a walk with given decomposition \eqref{walkDec}, and let $T,T'$ be gauge equivalent translation systems, so that there is a gauge transformation $V$ such that $T_\alpha'=VT_\alpha V^*$. Then the walks $\widetilde W, \widetilde W'$ arising by minimal coupling from $T,T'$, respectively, satisfy $\widetilde W'(t)=V(t)W(t)V(t)^*$ for all $t$.
\end{lem}

This kind of equivalence between $\widetilde W$ and $\widetilde W'$ may not always be obvious to see, but it has strong consequences. If we choose an initial state $\psi'(0)=V(0)\psi(0)$, and iterate the equation of motion \eqref{dynamics} with $\widetilde W'$ and $T_0'$ to achieve the state $\psi'(t)$, we can equivalently iterate $\widetilde W$ with $T_0$ and apply $V(t)$ at the end. In particular, the position probabilities will be the same and also the the internal degrees of freedom, conditioned on any position $x$. What will differ in general are matrix elements involving different positions.

We have seen that one of the directions in a translation system may be chosen to be the same as the flat system. It is natural to do this for the time direction. Then $T_0$ just identifies the Hilbert spaces $\HH(t)$, so we can work in a fixed Hilbert space with time translations given exclusively by the walk operator. We call this a \textbf{temporal gauge}. When the operator $T_0W(t)$ does not depend on time its iteration determines the long-time behaviour of the system. It is only in this case that spectral analysis is a helpful tool for studying the propagation behaviour. 
A discussion for the case of purely electric fields in terms of spectral properties is given in Example~\ref{sec:ewalks}.

\subsection{Uniqueness of minimal coupling}
The decomposition \eqref{walkDec} is not unique, so the natural question arises whether the result of minimal coupling depends on this choice. The answer is yes, and this is not an artefact of the discrete unitary setting. In fact it arises in almost the same way in the continuum setting.

Suppose we have some Hamiltonian $H$ given as a polynomial in the position and momentum operators.
Then replacing every momentum operator $P_\mu$ by $P_\mu-A_\mu(t,Q)$ as in \eqref{kinmom} gives another Hamiltonian $\widetilde H$,
interpreted as ``the same'' Hamiltonian placed in an external field described by the vector potential $A_\mu$.
The reason for repeating this description of the minimal coupling procedure as a straightforward substitution is to point out a hidden assumption:
The Hamiltonian must be presented as a polynomial and, in fact, as a non-commutative polynomial, in which monomials with different operator orderings are considered a priori as different. Without this preparatory step the substitution \eqref{kinmom} is just not defined\footnote{Polynomials might be a bit too narrow here if one thinks of functions such as $p\mapsto(p^2+m^2)^{1/2}$. So in general one would want to include the full non-commutative functional calculus (see, e.g., the Appendix of \cite{MF}). For bounded arguments this is covered by a Weiertra{\ss}-type approximation theorem, so polynomials do tell the essential part of the story. For unbounded arguments, as in the present case, the practice in physics is to use the scheme as a formal device, and to look at mathematical subtleties only for the cases one is really interested in, that is, usually later or never. One can easily cook up examples, where minimal substitution on a free particle leads to a Hamiltonian that is not essentially self-adjoint, so dubious as a generator of dynamics}.

Hence the substituted Hamiltonian $\widetilde H$ does not just depend on the {\it operator} $H$. Consider, for example,
the Hamiltonians
\begin{equation}\label{p1p2}
  H_1=P_1P_2=P_2P_1=H_2.
\end{equation}
The equality in the middle holds, because the ungauged momenta commute. After minimal coupling we get
\begin{equation}\label{p1p2coup}
  \widetilde H_1=\widetilde P_1\widetilde P_2\neq\widetilde P_2\widetilde P_1=\widetilde H_2.
\end{equation}
Now $\widetilde H_1-\widetilde H_2=F_{12}$ is the field, which will be non-zero in any non-trivial case. In the given context it seems possible that the two operators are equal up to a gauge transformation. But considering
\begin{eqnarray}
  e^{i\chi(Q)}\widetilde H_1e^{-i\chi(Q)}-\widetilde H_2&=&\\ &&\strut\mkern-160mu=
  \widetilde H_1-\widetilde H_2-(\partial_1\chi)\magtcont{2}-\magtcont{1}(\partial_2\chi)+(\partial_1\chi)(\partial_2\chi), \nonumber
\end{eqnarray}
we see that the left hand side can only vanish, when the coefficients of the differential operator on the right are zero. This still leaves us with $F_{12}=0$, so a gauge transformation does not help to achieve equality.
In other words, the minimal coupling procedure is not as straightforward as it is often presented in Quantum 101.

It is the same ambiguity that we encountered for unitary operators. Here, too, the difference cannot be covered by a gauge transformation. We can see this already for a minimal extension of the above example of pure shifts: consider decomposed walks on $\ell_2(\Ir^2)\otimes\Cx^2$ defined as
\begin{equation}
  W_1=S_1S_2=S_2S_1=W_2,
\end{equation}
where $S_i=P\Tsf_i+(1-P)\Tsff*_i$ and $P$ is a constant projection onto some one-dimensional subspace of $\Cx^2$.
As in the Hamiltonian example above, the equality in the middle holds because the flat translation operators $\Tsf_i$ commute. After discrete minimal coupling $\Tsf_i\mapsto\Ts_i$ we get

\begin{equation}
  \widetilde W_1=\begin{pmatrix}
    \Ts_1\Ts_2  &               \\
            &   \Ts_1^*\Ts_2^*
  \end{pmatrix}
  \neq
  \begin{pmatrix}
    \Ts_2\Ts_1  &               \\
            &   \Ts_2^*\Ts_1^*
  \end{pmatrix}
  =\widetilde W_2.
\end{equation}

Finding a gauge transformation $V$ which maps $\widetilde W_2$ to $\widetilde W_1$ amounts to finding solutions to
\begin{equation}
  V^*\widetilde W_2^*V\widetilde W_1\phi_x=\idty.
\end{equation}
This leads to the two conditions
\begin{align}
  V(x+\hat1+\hat2)  &= P_{12}^*(x)V(x),  \\
  V(x+\hat1+\hat2)  &= P_{12}(x)V(x),
\end{align}
which can be satisfied iff $P_{12}(x)=\pm\idty$ for all $x$. Thus, for arbitrary fields the walks $\widetilde W_1$ and $\widetilde W_2$ are not gauge equivalent, and minimal coupling always involves a choice of operator ordering.

\subsection{Discussion of minimal coupling}
The result of the minimal coupling is another dynamical system of the same kind. Indeed, if we have written $T$ as the flat translation system times $U_\alpha(x)$ as in \eqref{UaxG} a subshift and its substituted version differ by a unitary factor, which we can make part of an adjacent coin, so we are back to a decomposition with respect to a flat translation system. This shows that there is no absolute distinction between walks without or with field: The procedure of adding a field is relative. It can be iterated, but the result of several substitutions can be read off from what happens at the level of translation systems, and can just as well be done in a single step. This ``addition of fields'' is commutative in an abelian gauge theory.

The only reason why we have restricted to $U(1)$-gauge fields is the gauge equivalence expressed in Lemma~\ref{lem:gaugeeq}. The argument for this requirement depends crucially on the coins commuting with the gauge transformations, which is automatic for a $U(1)$-theory. There seem to be three possibilities to deal with this: One could either decide to live with the failure of that property, or (preferably) demand that the coins commute with $\ggroup$, or include a transformation of the coins in the definition of minimal coupling. In the spirit of this paper, the decision should be inspired by a study of non-abelian continuum theories, but that would be beyond the scope of our paper.

\section{Examples}
\subsection{Homogeneous Systems}\label{sec:homo} 

The idea of a homogeneous system is clear enough: Its properties should be everywhere the same. More formally, the translations should act as a symmetry group. Quantum systems in a constant external magnetic field are a notorious example showing that this does not imply that a description in terms of translation invariant quantities is possible. Indeed, the vector potential for a non-zero constant field necessarily grows at least linearly in the coordinates. The translations in this case are indeed a symmetry, but only up to gauge transformations. So it is natural to introduce another translation system $S$, which unites the required shifts and gauge transformations. Indeed, the notion of translation system is ideally suited to express such combinations. Given the translation system $T$ expressing the electromagnetic field, the condition that $S$ acts as a symmetry is simply
\begin{equation}\label{stts}
  S_\alpha T_\beta= T_\beta S_\alpha
\end{equation}
for all $\alpha,\beta$. Note that $T_\alpha$ cannot express the symmetry operations, because for non-vanishing field the component translations do not commute, see \eqref{eq:plaquetteoperator}.

Let us look at the continuous case for guidance to the right questions. The analogy starts from a connection $\partial_\alpha-iA_\alpha$ (infinitesimal version of $T_\alpha$), and asks for the existence of a second connection, $\partial_\beta-iB_\beta$, expressing the symmetry, in the sense of commuting with the first. The condition for that is obviously
\begin{equation}\label{dBdAdAdB}
  \partial_\alpha B_\beta=\partial_\beta A_\alpha
\end{equation}
for all $\alpha,\beta$. The first consequence is that $\partial_\alpha(A_\beta+B_\beta)=\partial_\beta(A_\alpha+B_\alpha)$. That is, the curvatures (=fields) add up to zero, which implies that
$A_\alpha+B_\alpha=\partial_\alpha C$ for some scalar function $C$. This turns \eqref{dBdAdAdB} into an equivalent equation for $C$, namely
\begin{equation}\label{dabC}
  \partial_\alpha\partial_\beta C=\partial_\beta A_\alpha+\partial_\alpha A_\beta.
\end{equation}
$C=0$ solves this equation if the right hand side vanishes, i.e., $A$ satisfies the \textbf{symmetric gauge} condition, which is quite popular for constant magnetic fields. It actually does not exist otherwise, as one sees from differentiating \eqref{dabC} with respect to $x_\mu$ and using the permutation symmetry of $\partial_\mu\partial_\alpha\partial_\beta C$. This readily gives $\partial_\mu F_{\alpha\beta}=0$, i.e.,  the fields are constant, which is satisfying, because it implies that also the $\partial_\beta-iB_\beta$ commute up to constants. Hence their exponentials, which are the unitary operators expressing the symmetry of finite translations commute up to global phases. This is just what Wigner's theorem requires of a symmetry. So we would have had to impose this condition anyway, but it turns out it came out just from \eqref{stts}. In the literature these unitaries are known as \textbf{magnetic translation operators} \cite{brown1964bloch,zak1,zak2}.

The discrete analogue of these statements is the following.

\begin{prop}Let $T$ be an $U(1)$-translation system. Then the following conditions are equivalent:
\begin{itemize}
\item[(1)] There is a second translation system $S$ such that $S_\alpha T_\beta= T_\beta S_\alpha$.
\item[(2)] The plaquette phases $P_{\alpha\beta}$ are independent of $x$.
\item[(3)] $T$ is gauge equivalent to a translation system $T^\prime$ that satisfies $T_\alpha^\prime\phi_x =\Tf_\alpha P^>_{\alpha}(x)\phi_x$ with $P^>_\alpha(x) = \prod_{\beta>\alpha} (P_{\alpha\beta})^{x_\beta}$, see Fig.~\ref{fig:pathorder}.
\end{itemize}
In this case $S$ in (1) is defined up to a constant phase, has plaquette phases $1/P_{\alpha\beta}$, and in the gauge (3) takes the form $S_\alpha^\prime \phi_x =  \Tf_\alpha P^<_\alpha(x)\phi_x$, with $P^<_\alpha(x)=\prod_{\gamma<\alpha}(P_{\gamma\alpha})^{x_\gamma}$.
\end{prop}

\begin{proof}
\proofdir12
According to \eqref{eq:plaquetteoperator}, $P_{\alpha\beta}$ is given as the product $T_\alpha^*T_\beta^*T_\alpha T_\beta$ and hence clearly invariant under conjugation by $S_\gamma$ if we assume (1). At the same time, $P_{\alpha\beta}$ is a localized operator on which the translation system $S_\gamma$ acts as $S_\gamma P_{\alpha\beta}S^*_\gamma\phi_x = P_{\alpha\beta}(x-\hat{\gamma})\phi_x$, hence $P_{\alpha\beta}(x)$ cannot depend on $x$.

\proofdir23 Choosing the path-ordered gauge according to \eqref{pathorder} and using repeatedly $T^\prime_\alpha T^\prime_\beta = P_{\alpha\beta}T^\prime_\beta T^\prime_\alpha$, we find
\begin{align}
  T^\prime_\alpha \phi_x &= T^\prime_\alpha \left(T^\prime_s\right)^{x_s}\cdots \left(T^\prime_\alpha\right)^{x_\alpha}\cdots \left(T^\prime_0\right)^{x_0} \phi_0\\
  &= \left(\prod_{\beta>\alpha} \left(P_{\alpha\beta}\right)^{x_\beta}\right)\left(T^\prime_s\right)^{x_s}\cdots \left(T^\prime_\alpha\right)^{x_\alpha+1}\cdots \left(T^\prime_0\right)^{x_0}\phi_0\,.
\end{align}

\proofdir31 Choosing the path ordered gauge and setting  $S^\prime_\alpha \phi_x =  P^<_\alpha(x) \Tf_\alpha\phi_x $, a direct calculation using \eqref{plaqx} shows that both $S^*_\alpha S^*_\beta S_\alpha S_\beta = 1/P_{\alpha\beta}$ and $S_\alpha T_\beta= T_\beta S_\alpha$ are satisfied.
\end{proof}

The $T'_\alpha$ in this proposition are a basis for a symmetry representation of the translation group in the following sense: for each $x\in\Ir^s$ let $\gamma(x)$ be the standard path $0\to x$. Then $x\mapsto T'_{\gamma(x)}$ corresponds to a projective representation with
\begin{equation}
  T'_{\gamma(x)}T'_{\gamma(y)}=P^>(x,y)T'_{\gamma(x+y)},
\end{equation}
where $P^>(x,y)=\prod_{\beta>\alpha}P_{\alpha\beta}^{x_\alpha y_\beta}$ is a multiplier \cite{BackhouseProjective1,BackhouseProjective2}. It follows from standard arguments that $T'_{\gamma(x)}$ commutes with the projective representation with multiplier $P^<(y,x)$ up to normalization of $P^>(x,y)$ \cite{kleppner1962,kleppner1965}.

\subsection{Rational fields}\label{sec:ratfields}
In general, even under the homogeneity assumption electromagnetic systems are not easy to solve. The main reason is that Fourier transformation fails. After all, the idea is to jointly diagonalize translations and the walk or Hamiltonian, which fails, when the translations do not commute in the first place. However, translation invariance can sometimes be restored for a magnetic system by \textbf{regrouping}. For example, in the two dimensional case of Sect.~\ref{sec:butter}, when $F_{12}=p/q$ is rational the translation in $x$-direction commutes with the translation by $q$ steps in $y$-direction. Hence, if we group cells periodically to supercells of $q$ individual cells stacked in $y$-direction, we come back to a strictly translation invariant system, which can be solved by Fourier transform. The internal structure of the supercells now means that we have $q$ times the number of internal degrees of freedoms, and correspondingly many bands. It is clear even in this simplest example that the regrouping is not unique, and this will persist in more complex cases.

The following proposition describes the kind of system for which regrouping to a translation invariant system is feasible. Note that ``rational'' phases are those for which some power is $1$, so that in the present context we call $F$ \textbf{rational} if $F\in2\pi\Rt$.

\begin{prop}
Let $T$ be a homogeneous translation system on $\Ir^s$, $s\geq2$ with field matrix $F$. Then the following are equivalent
\begin{itemize}
\item[(1)] All entries of $F$ are rational, i.e., $F_{\mu\nu}\in2\pi\Rt$ for all $\mu,\nu$.
\item[(2)] There is a sublattice $\Lambda\subset\Ir^s$, generated by linearly independent vectors $\lambda_1,\ldots,\lambda_s$ such that the translations by lattice vectors commute.
\item[(3)] The holonomy group $\hol$ is finite.
\end{itemize}
Each of these conditions comes with a natural size parameter, namely
\begin{eqnarray}
  q_1 &=&\min\{q\in\Nl|\forall\mu,\nu:\  qF_{\mu\nu}\in2\pi\Ir\} \\
  q_2 &=&\min|\det(\lambda_1,\ldots,\lambda_s)|\\
  q_3 &=& \#\hol=q_1
\end{eqnarray}
\end{prop}

The number of practical interest here is $q_2$, which is the number of lattice points of $\Ir^s$ lying in an elementary cell of the lattice $\Lambda$.

\begin{proof}
In the proof we will keep track of the numbers $q_i$, providing some basic bounds.

\noindent{\it (1)$\Leftrightarrow$(3):\ }
Since $P_{\alpha\beta}=\exp(iF_{\alpha\beta})\in\hol$. If this group is finite, it consists of the $q_3^{\rm th}$ roots of unity, so $q_3F_{\alpha\beta}\in2\pi\Ir$. Conversely, if all  $F_{\alpha\beta}$ are rational with denominator $q_1$ the plaquette phases are all in the group of $q_1^{\rm th}$ roots of unity, and since all closed paths can be composed of plaquettes, this must be the whole holonomy group. Hence $q_1=q_3$.

\proofdir12
Take $\lambda_\alpha=q_1\,\hat\alpha$ for $\alpha<s$ and $\lambda_s=\hat s$. Then $F(\lambda_i,\lambda_j)\in2\pi\Ir$, because at least one of the vectors involved has a factor $q_1$. Hence $q_2\leq q_1^{s-1}$.
Of course, there may be smaller lattices with this property.

\proofdir21
Let $\Lambda$ be a lattice of commuting translations. Consider the dual lattice $\Lambda'\subset\Rl^s$, which is spanned by a dual basis, i.e., vectors $\xi_k$ so that $\xi_k\cdot\lambda_j=\delta_{ik}$. The matrix of components of the $\xi_k$ is the inverse of the component matrix of the $\lambda_j$, whose determinant is $q_2$. Hence the components of $\xi_k$ are rational with denominator $q_2$. Expressing the basis vectors of $\Ir^s$ in the basis $\{\lambda_i\}$, and observing that $F$ takes $2\pi\Ir$-values on pairs of such basis vectors, we get that $F$ is rational with denominator $q_2^2$. That is $q_1\leq q_2^2$.
\end{proof}

The one-dimensional case was excluded here, because holonomy is trivial and there are no plaquette phases. In the two-dimensional case there is only one field component, say, $B=F_{12}=2\pi p/q$. Then for any pair of integer vectors $x,y$ we have
\begin{equation}\label{2Dregroup}
  \sum_{\alpha\beta}F_{\alpha\beta}x_\alpha\,y_\beta=2\pi\frac pq\ (x_1y_2-x_2y_1) =2\pi\frac pq\ \det(x,y).
\end{equation}
Hence $x$ and $y$ qualify as basis vectors $\lambda_1,\lambda_2$ satisfying (2) iff their determinant is a multiple of the denominator $q$. Clearly, the minimal choice is $q$ itself, so that $q_1=q_2=q_3$ in this case.
Preliminary checks suggest that this might also be the case for $s=3$, but for higher dimension we do not yet have a convincing intuition.

\subsection{Electric walk 1D}\label{sec:ewalks}

Let us briefly come back to the purely electric setting introduced in Sec. \ref{sec:mincoup}  $P_{k\ell}=\idty$ for $k,\ell=1,\dots,s$ and $P_{0\alpha}$ is the electric field component in direction $\alpha$. Indeed, in the gauge where the spatial translations are flat, ``switching on'' an electric field by $\Tf_0\mapsto\Tf_0 U_0(t,x)$ boils down to \cite{ewalks}
\begin{equation}\label{eq:elecFSw}
  W\mapsto U_0(t,x)W.
\end{equation}
Note however, that this procedure corresponds to a particular choice of gauge. Other common gauges are the temporal (or ``Weyl'') gauge, were $\Tf_0$ is left unchanged \cite{creutz1977gauge}, and in the homogeneous case the symmetric gauge.

In one spatial dimension quantum walks in homogeneous and static electric fields have been studied extensively, both theoretically \cite{ewalks} and experimentally \cite{meschede13}. There, electric walks are described in the gauge \eqref{eq:elecFSw}, in which the electric walk operators become time-independent, which allows to meaningfully discuss their spectral properties.

For electric walks of the form $W=e^{i\EF Q}CS$ the propagation behaviour depends discontinuously on the field $\EF$ \cite{ewalks} or, more precisely on the rationality or irrationality of $E/(2\pi)$. In the rational case, one can regroup local cells as described in Sec. \ref{sec:ratfields} in order to obtain a translation invariant quantum walk with a larger internal degree of freedom. This eventually leads to ballistic expansion, whereas on short time scales of the order of the denominator of the field revivals to the initial state are found. In contrast, irrational fields lead either to Anderson localization \cite{locQuasiPer} similar to the disordered setting \cite{dynloc,dynlocalain}, or they propagate hierarchically. In the latter case, the particle shows an infinite number of sharper and sharper revivals of the initial state. After each of these revivals the particle propagates farther and farther. Each of these propagation behaviours corresponds to a different spectral type: in the rational case the ballistic expansion corresponds to absolutely continuous spectrum, whereas localization in the irrational case is characterized by pure point spectrum. For irrational fields which are enormously well approximable by rationals the spectrum is singular continuous.

\subsection{Walk with quasi-periodic coin}\label{sec:qperiodic}

In temporal gauge $U_0=\idty$ the spatial translations cannot be chosen flat anymore, and the electric walks in the previous section become explicitly time-dependent. For the one-dimensional electric walk of \cite{ewalks} this gauge transformed walk operator was shown in \cite{ususi} to be given by $W(t)=CS(t)$, where $S(t)=\sum_{k=\pm1} P_k\Ts_1^k$ for $\Ts_1=\Tsf_1U_1(x,t)$ with $U_1(x,t)=e^{-itE}$ and $P_k$ denotes the projection onto the $k$-eigenstate of $\sigma_z$.

In \cite{susi}, a similar model with quasi-periodically time-dependent coin $C(t)=R_y(\theta)R_x(t\phi)$ was discussed, where $R_\alpha$ denotes rotation around the $\alpha$-axis in coin space. Even though the translation systems in this walk model and the electric walk are not $U(1)$-equivalent but only equivalent up to a Hadamard coin, their propagation behaviour is strikingly similar. In particular, for the walk $W(t)=C(t)S$ the same revival structure is observed. As discussed in \cite{ususi}, the reason for this is that the same techniques apply which, however, is a special feature of this model.

\subsection{Magnetic walk 2D}\label{sec:butter}  

To give a concrete example of the purely magnetic case, let us consider the simplest setting in which magnetic fields can occur, i.e. a two-dimensional lattice \cite{Sajid,marquez2018electromagnetic}. On $\ell_2(\Ir^2)\otimes\Cx^2$ we consider the decomposed walk
\begin{equation}
  W=S_2CS_1C'
\end{equation}
where $C,C'$ are localized unitaries. The $S_\alpha$ are state-dependent shifts defined by $S_\alpha:=\sum_{k=\pm1} P_k\Tsff{k}_\alpha$ where as above $P_k$ denotes the projection onto the $k$-eigenstate of $\sigma_z$.

Since there are no electric fields present we have $P_{0\alpha}=\idty$ throughout and the magnetic fields are static by the Maxwell equations \eqref{eq:disc_maxwell}. Magnetic fields identified by non-trivial spatial plaquette phases are implemented via the minimal coupling $\Tsf_\alpha\mapsto \Ts_\alpha$ for $\alpha=0,1,2$. As discussed at the end of Sec.~\ref{sec:mincoup}, in the purely magnetic case it is advantageous to work in temporal gauge: in this (partial) gauge the $U_k(t,x)$ are time-independent and consequently dynamics are implemented by iterating the same walk operator.

\begin{figure}[t]
\begin{center}
  \includegraphics[width=.4\textwidth]{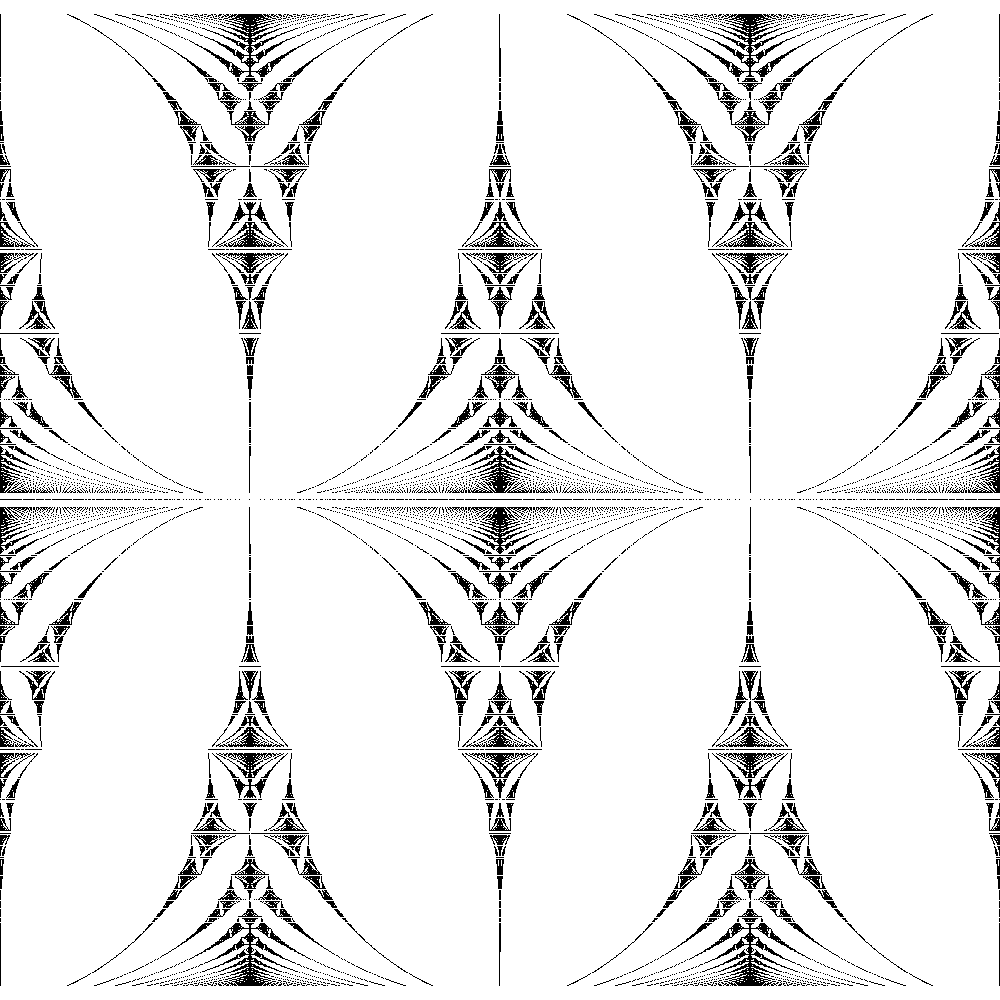}
  \caption{\label{Butterfliege}The spectrum of a two-dimensional magnetic Hadamard walk \cite{christalk}. The vertical axis corresponds to the field and the horizontal axis to the argument of the quasi-energy.}
\end{center}
\end{figure}

Assuming the magnetic field $F_{12}$ to be homogeneous and the coin operators $C,C'$ to be given as the constant Hadamard matrix $\left(\begin{smallmatrix} 1&\hphantom{-}1\\1&-1\end{smallmatrix}\right)/\sqrt2$, the spectrum of the magnetic walk in dependence of the field resembles a fractal heavily reminding of the famous \emph{Hofstadter butterfly} \cite{hof76}, see Fig.~\ref{Butterfliege}. The symmetric structure of this ``Quantum Walk Butterfly'' can be understood using techniques from irrational rotation algebras which, however, is beyond the scope of this example.

For $F_{12}/(2\pi)=\nr/\dr$ rational the system is translation invariant after regrouping, see Sec.~\ref{sec:ratfields}. Therefore, the spectrum is absolutely continuous and consists of $2\dr$ bands. For irrational fields the spectrum is expected to be homeomorphic to the Cantor set, similar to the continuous system of Bloch electrons in a magnetic field \cite{bellissard1982cantor}\cite[Chapter VI.]{hof76}.
A similarly discontinuous dependence of the spectral type on the field parameter was observed for one-dimensional electric walks \cite{ewalks}, see also Sec.~\ref{sec:ewalks}, and for the original butterfly \cite{hof76}. However, since the distinction between the different classes of fields requires infinite precision, it remains unclear how it influences real-life experiments.

\appendix
\section{From continuous to discrete - and back}\label{app:DisCont}

Let us here lay out the details of the connection between the differential calculus on smooth manifolds and the discrete differential calculus on $\Ir^s$. This will allow us to conclude a Poincar\'e lemma on $\Ir^s$ from that on $\Rl^s$.

On both sides we deal with expressions which are sums of terms of the form $f(x)\dd x^{\alpha_1}\wedge \dd x^{\alpha_1}\cdots \wedge\dd x^{\alpha_p}$. The coefficient function $f$ will be a function of position, i.e., $x\in\Rl^s$ in the continuum case and $x\in\Ir^s$ in the discrete case. Its values will be in $\Rl$ in the continuum case, and in the gauge group $U(1)$ in the discrete case. Both abelian groups will be written additively, so that, although we really mean products in the group of phases, we write sums of terms in $\Rl/2\pi\Ir$. We do not require here any multiplication of these coefficients, although, of course, this is well-defined in the continuum case, and is needed to define the wedge product of differential forms. The coordinate differentials ``$\dd x^{\alpha}$'' are used as a purely formal device to aid the bookkeeping of antisymmetric expressions. Their wedge products are completely defined by being associative and antisymmetric.
Hence any product of differentials can be brought into the form
\begin{equation}\label{dxI}
  \dd x^{\alpha_1}\wedge \dd x^{\alpha_1}\cdots \wedge\dd x^{\alpha_p}=:\dd x^I
\end{equation}
where $I=\{\alpha_1,\alpha_2,\ldots,\alpha_p\}\subset\{1,\ldots,s\}$ with $\alpha_1<\alpha_2<\cdots\alpha_p$. The ordering process of a similar expression with permuted $\alpha_i$ to the normal form \eqref{dxI} at most produces signs, which makes sense in the respective coefficient group. The group of forms will be denoted in the continuum case by
\begin{equation}
  \mathcal C=\bigwedge^s\nolimits\mathcal C(\Rl^s; \Rl) \end{equation}
with coefficients in $\mathcal C(\Rl^s;\Rl)$ the space of suitably differentiable functions $f:\Rl^s\to\Rl$. The degree of differentiability will be indicated in the context. Similarly, we write in the discrete case
\begin{equation}
  \mathcal D=\bigwedge^s\nolimits\mathcal C(\Ir^s;U(1))
\end{equation}
where the coefficients are taken as elements of $\mathcal C(\Ir^s;U(1))$ the space of functions from the lattice to the additively written group of phases, i.e., $\Rl/2\pi\Ir$.

Exterior derivatives $d$ taking $p$-forms to $p+1$-forms are defined in the smooth case as
\begin{equation}
  d\bigl(f\,\dd x^I\bigr)=\sum_{\alpha\in I^c}(\partial_\alpha f)\, \dd x^\alpha\wedge \dd x^I,
\end{equation}
where $\partial_\alpha f=\partial f/\partial x^\alpha$, and $I^c$ denotes the complement of $I$ in $\{1,\ldots,s\}$. In the discrete case we write
\begin{equation}\label{eq:ddisc}
  d\bigl(f\,\dd x^I\bigr)=\sum_{\alpha\in I^c}(d_\alpha f)\, \dd x^\alpha\wedge \dd x^I,
\end{equation}
where $(d_\alpha f)(x)=f(x+\alpha)-f(x)$ denotes the discrete derivative in direction $\alpha$. Clearly, both kinds of partial derivatives commute, which implies the fundamental relation
\begin{equation}\label{eq:d20}
  d^2=0.
\end{equation}
In either setting, forms $f$ which satisfy $df=0$ are called \textbf{closed} whereas forms $f$ for which there exists another form $g$ such that $f=dg$ are called \textbf{exact}. It follows immediately from \eqref{eq:d20} that exact forms are closed. The converse is known as the

\begin{lem}[Poincar\'e Lemma]\label{lem:poincont}
  Every closed form is exact.
\end{lem}

For $\mathcal C$ this is well known \cite{LeeDiffGeo}, and can be shown also for subregions, provided they are ``star-shaped''. For the discrete case this is done in \cite{hydon2004variational,mansfield2008difference}, with a subtle discussion of what star shaped should mean in the discrete case. We will not need this, but only the global version for the entire lattice. In order to strengthen the connections between the two calculi we sketch a proof by which the discrete result is derived from the smooth one. This may not be the natural order (as the discrete result is in some sense more elementary), but we hope it reduces the less familiar to the more familiar for most of our readers.

{\bf Discretization} was already discussed in Sect.~\ref{sec:tight}. Here we get it as a map $\Delta:\mathcal C\to\mathcal D$. Its counterpart is  a {\bf continuization}, a  map $\Gamma:\mathcal D\to\mathcal C$. We will show that
\begin{equation}\label{eq:DGid}
  \Delta\Gamma=\textrm{id},
\end{equation}
i.e., first continuizing and then discretizing again gets us back to where we started from. The opposite relation is bound to fail, because discretization is clearly many-to one.

To find such maps we first define in one dimension $\Delta_0,\Delta_1:\mathcal C(\Rl)\to\mathcal C(\Ir)$ by
\begin{align}\label{eq:conttodisc}
  (\Delta_0f)(n)    &=f(n)   \\
  (\Delta_1f)(n)    &=\int_n^{n+1} dx f(x).
\end{align}
Conversely,
$\Gamma_0,\Gamma_1:\mathcal C(\Ir)\to\mathcal C(\Rl)$ are defined by
\begin{align}
	  (\Gamma_0f)(x)    &=(x-\lfloor x\rfloor)f(\lceil x\rceil)+(\lceil x\rceil-x)f(\lfloor x\rfloor)\\[1mm]
  (\Gamma_1f)(x)    &=f(\lfloor x\rfloor)\label{eq:disctocont}.
\end{align}
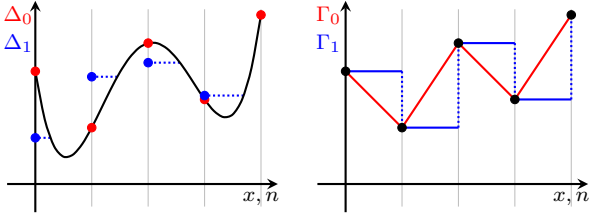
\begin{figure}[t]
	\begin{center}
%
%

\tikzset{
	>=stealth,
	right graphic/.style={
		xshift=5.5cm
	}
}

\begin{tikzpicture}[font=\footnotesize,scale=.75]
\def\dxx{4.3};
\def\dyy{3.25};
\def\drr{.75mm};
\def\pzero{(0,2)};
\def\pone{(1,1)};
\def\ptwo{(2,2.5)}
\def\pthree{(3,1.5)};
\def\pfour{(4,3)};

\foreach \n in {1,2,3,4} {
	\draw[lightgray] (\n,-.4) -- (\n,3.1);
};
\draw[black,thick,->] (-.5,0) -- (\dxx,0);
\draw[black,thick,->] (0,-.5) -- (0,\dyy);

\node (labelx) at (4,-.25) {$x,n$};
\node (labely1) at (-.3,3) {\textcolor{red}{$\Delta_0$}};
\node (labely2) at (-.3,2.5) {\textcolor{blue}{$\Delta_1$}};

\draw[blue,thick,dash pattern=on \pgflinewidth off 1pt] (1.4535,1.90278) -- (1,1.90278);
\draw[blue,thick,dash pattern=on \pgflinewidth off 1pt] (2.56417,2.15278) -- (2,2.15278);
\draw[blue,thick,dash pattern=on \pgflinewidth off 1pt] (3.69307,1.56944) -- (3,1.56944);
\draw[blue,thick,dash pattern=on \pgflinewidth off 1pt] (0.267019,0.819444) -- (0,0.819444);

\draw[domain=0:4,smooth,variable=\x,black,thick] plot ({\x},{2-6.41666*\x+8.33333*\x*\x-3.33333*\x*\x*\x+0.4166666*\x*\x*\x*\x});

\draw[red,fill] \pzero circle (\drr);
\draw[red,fill] \pone  circle (\drr);
\draw[red,fill] \ptwo circle (\drr);
\draw[red,fill] \pthree circle (\drr);
\draw[red,fill] \pfour circle (\drr);

\draw[blue,fill] (1,1.90278) circle (\drr);
\draw[blue,fill] (2,2.15278) circle (\drr);
\draw[blue,fill] (3,1.56944) circle (\drr);
\draw[blue,fill] (0,0.819444) circle (\drr);

\foreach \n in {1,2,3,4} {
	\draw[right graphic,lightgray] (\n,-.4) -- (\n,3.1);
};
\draw[right graphic,black,thick,->] (-.5,0) -- (\dxx,0);
\draw[right graphic,black,thick,->] (0,-.5) -- (0,\dyy);

\draw[right graphic,red,thick] \pzero -- \pone -- \ptwo -- \pthree -- \pfour;
\draw[right graphic,blue,thick] \pzero -- +(1,0);
\draw[right graphic,blue,thick] \pone -- +(1,0);
\draw[right graphic,blue,thick] \ptwo -- +(1,0);
\draw[right graphic,blue,thick] \pthree -- +(1,0);
\draw[right graphic,blue,thick,dash pattern=on \pgflinewidth off 1pt] (1,2) -- \pone;
\draw[right graphic,blue,thick,dash pattern=on \pgflinewidth off 1pt] (2,1) -- \ptwo;
\draw[right graphic,blue,thick,dash pattern=on \pgflinewidth off 1pt] (3,2.5) -- \pthree;
\draw[right graphic,blue,thick,dash pattern=on \pgflinewidth off 1pt] (4,1.5) -- \pfour;

\draw[right graphic,black,fill] \pzero  circle (\drr);
\draw[right graphic,black,fill] \pone  circle (\drr);
\draw[right graphic,black,fill] \ptwo circle (\drr);
\draw[right graphic,black,fill] \pthree circle (\drr);
\draw[right graphic,black,fill] \pfour  circle (\drr);

\node (labelx) at (9.5,-.25) {$x,n$};
\node (labely1) at (5.2,3) {\textcolor{red}{$\Gamma_0$}};
\node (labely2) at (5.2,2.5) {\textcolor{blue}{$\Gamma_1$}};
\end{tikzpicture}

	\end{center}
	\caption{\label{fig:DeltaGamma} Action of the discretization (left) and continuization (right) maps $\Delta_i$ and $\Gamma_i$  \eqref{eq:conttodisc}-\eqref{eq:disctocont}.}
\end{figure}
Here the floor function is defined as $\lfloor x\rfloor=\max\{n\in\Ir|n\leq x\}$, and the ceiling function (non-standardly) as $\lceil x\rceil=\lfloor x\rfloor+1$.  Hence, for an integer $n\in\Ir$,
$\lfloor n\rfloor=n$ and $\lceil n\rceil=n+1$. Thus both functions are lower semicontinuous. Fig.~\ref{fig:DeltaGamma} shows these functions. Obviously, the ranges of $\Gamma_0$ and $\Gamma_1$ do not consist of smooth functions, however, as piecewise continuous functions they do make sense as integrands of forms over arbitrary bounded regions. Moreover, for the purpose of such integrals, they can be approximated pointwise by smooth functions, making the integrals converge by dominated convergence. Henceforth we replace $\mathcal C(\Rl;\Rl)$ by the algebra of piecewise continuous, lower semicontinuous, and locally bounded functions.
One quickly verifies that
\begin{equation}\label{eq:D01G01id}
  \Delta_0\Gamma_0=\textrm{id}=\Delta_1\Gamma_1.
\end{equation}
(Actually, also $\Delta_0\Gamma_1=\textrm{id}$, but $\Delta_1\Gamma_0\neq \textrm{id}$, but these are not needed).
If we take the $\Delta_p$ and $\Gamma_p$ to act on the coefficients of $p$-forms for $p=0,1$, respectively, we immediately verify that
\begin{equation}
  d\Delta_0=\Delta_1d
\end{equation}
as well as
\begin{equation}
  d\Gamma_0=\Gamma_1d,
\end{equation}
where it is understood from the context which of the two exterior derivatives is meant by $d$.

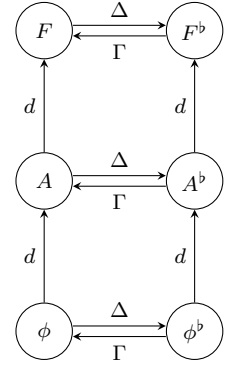
\begin{figure}[t]
\begin{center}
%


\tikzstyle{block} = [draw, circle, radius = .4, minimum width = 0.75cm, minimum height = 0.75cm]
\tikzstyle{input} = [coordinate]
\tikzstyle{output} = [coordinate]

\begin{tikzpicture}
	[
	font=\footnotesize,
	node distance=2cm,auto,>=latex'
		]

    \node [block,name=F] {$F$};
    \node [block,below of=F] (A) {$A$};
    \node [block,below of=A] (phi) {$\phi$};

    \node [block,name=Ff,right of=F] {$F^\flat$};
    \node [block,below of=Ff] (Af) {$A^\flat$};
    \node [block,below of=Af] (phif) {$\phi^\flat$};




    \draw[->,>=stealth] (A) to node {$d$} (F);
    \draw[->,>=stealth] (phi) to node {$d$} (A);
    \draw[->,>=stealth] (Af) to node {$d$} (Ff);
    \draw[->,>=stealth] (phif) to node {$d$} (Af);

    \draw[->,>=stealth] ([yshift=2]F.east) to node {$\Delta$} ([yshift=2]Ff.west);
    \draw[->,>=stealth] ([yshift=2]A.east) to node {$\Delta$} ([yshift=2]Af.west);
    \draw[->,>=stealth] ([yshift=2]phi.east) to node {$\Delta$} ([yshift=2]phif.west);

    \draw[<-,>=stealth] ([yshift=-2]F.east) to node[below] {$\Gamma$} ([yshift=-2]Ff.west);
    \draw[<-,>=stealth] ([yshift=-2]A.east) to node[below] {$\Gamma$} ([yshift=-2]Af.west);
    \draw[<-,>=stealth] ([yshift=-2]phi.east) to node[below] {$\Gamma$} ([yshift=-2]phif.west);

  \end{tikzpicture}
\end{center}
\caption{\label{fig:disc_diff_cal}Discretization and continuization of $0,1,2$-forms. This is commutative except that $\Gamma\Delta\neq\id$.}
\end{figure}

These maps allow us now to define $\Delta$ and $\Gamma$ for arbitrary lattice  dimension $s$ as follows: First, for any multi-index $I$ we define a map which discretizes elements of $\mathcal C(\Rl^s)$, i.e. $\Delta^I:\mathcal C(\Rl^s)\to\mathcal C(\Ir^s)$, by
\begin{equation}
  \Delta^I=\Delta_1^{\otimes I}\otimes\Delta_0^{\otimes I^c},
\end{equation}
where $I^c$ is the complement of $I$. This discretization can be extended to the algebra of $p$-forms with coefficients in $\mathcal C(\Rl^s)$ simply by defining
\begin{equation}
  \Delta f=\Delta({\sum_I}'f_I\dd x^I)={\sum_I}'(\Delta^If_I)\dd x^I.
\end{equation}
Analogously, we define the maps $\Gamma^I$ and $\Gamma$ which continuitize $f\in\mathcal C(\Ir^s)$ and $f\in\mathcal D$, respectively. From \eqref{eq:D01G01id} it immediately follows that $\Delta$ and $\Gamma$ satisfy \eqref{eq:DGid}. Moreover,
\begin{equation} \label{d-intertwine}
  \Delta d=d\Delta\quad \text{and}\quad   \Gamma d=d\Gamma.
\end{equation}
These relations are summarized in the commutative diagram Fig.~\ref{fig:disc_diff_cal}, for $0,1,2$-forms. Apart from emphasizing the close relations between the two calculi, this provides a way to import the Poincar{\'e} Lemma from the continuous to the discrete case.

\begin{proof}[Proof sketch of the discrete Poincar\'e Lemma]
The proof is basically an immediate consequence of the above construction. However, in this sketch we gloss over the question how much smoothness is needed for the continuum version to hold.

Let $f\in\mathcal D$ be a closed discrete form, i.e. $df=0$. Then, we construct a corresponding form $f'\in\mathcal C$ by setting $f'=\Gamma f$ which is closed by \eqref{d-intertwine}. According to the Poincar\'e lemma \ref{lem:poincont} $f'$ is exact, i.e. there exists a form $g'$ such that $f'=dg'$. Acting with $\Delta$ on both sides by \eqref{eq:DGid} gives
  \begin{equation}
    f=\Delta f'=\Delta dg'=d\Delta g'=dg,
  \end{equation}
  where we abbreviated $\Delta g'=g$.
\end{proof}

\vfill

\section*{Acknowledgements}
C. Cedzich acknowledges support by the Excellence Initiative of the German Federal and State Governments (ZUK 81) and the DFG (project B01 of CRC 183).

T. Geib and R. F. Werner acknowledge support from the ERC grant DQSIM, the DFG SFB 1227 DQmat, and the European project SIQS.

A. H. Werner thanks the Humboldt Foundation for its support with a Feodor Lynen Fellowship and the VILLUM FONDEN via the QMATH Centre of Excellence (Grant No. 10059).

\bibliography{mwalkshortbib}

\end{document}